\documentclass[fleqn,onecolumn,11pt]{wlscirep}
\usepackage{bm}
\usepackage{color}
\usepackage{color}
\usepackage{alltt,dsfont}
\usepackage{appendix}
\usepackage{amsmath,amssymb}
\usepackage{braket}
\usepackage[capitalise]{cleveref}
\usepackage{ulem}
\usepackage{float}
\usepackage{graphicx,subfigure,bbm}

\usepackage{lipsum}
\usepackage{nameref}
\usepackage{hyperref}
\usepackage{cleveref}

\usepackage[export]{adjustbox}
\usepackage{subscript}
\usepackage{titlesec,setspace}

\usepackage{caption}
\usepackage{float} 

\newfloat{extfigure}{htbp}{loe}
\floatname{extfigure}{Extended Data Fig.}

\newcommand{\onlinecite}[1]{\hspace{-1 ex} \nocite{#1}\citenum{#1}}

\newcommand{\refdisp}[1]{Ref. [\onlinecite{#1}]}
\newcommand{\figdisp}[1]{Fig. \ref{#1}}

\newcommand{\disp}[1]{Eq. (\ref{#1})}

\newcommand{\beq}{\begin{eqnarray}}
\newcommand{\eeq}{\end{eqnarray}}

\title{Twisting the Hubbard model into the Momentum-Mixing Hatsugai-Kohmoto Model}

\author[1,*]{Peizhi Mai}
\author[1]{Jinchao Zhao}
\author[1]{Gaurav Tenkila}
\author[1]{Nico A. Hackner}
\author[1]{Dhruv Kush}
\author[1]{Derek Pan}
\author[1,$\dagger$]{Philip W. Phillips}

\affil[1]{Department of Physics and the Anthony J. Leggett Institute of Condensed Matter Theory, University of Illinois at Urbana-Champaign, Urbana, IL 61801, USA\looseness=-1}

\affil[*]{peizhimai@gmail.com}          
\affil[$\dagger$]{dimer@illinois.edu}

\begin{abstract}

The Hubbard model is a standard theoretical tool for studying materials with strong electron-electron interactions, such as the cuprate superconductors. Unfortunately, interaction-driven phenomena such as the transition into the strongly correlated Mott insulator phase are difficult to treat with established theoretical techniques. However,  the exactly solvable Hatsugai-Kohmoto model displays similar Mott physics. Here we show how the Hatsugai-Kohmoto model can be deformed continuously into the Hubbard model. The trick is to systematically re-introduce all the momentum mixing the original Hatsugai-Kohmoto model omits. This can be accomplished by grouping $n$-momenta into a cell and hybridizing them resulting in the momentum-mixing Hatsugai-Kohmoto (MMHK) model.  We recover the Bethe ansatz ground state energy of the one-dimensional Hubbard model to within 1$\%$ from only ten mixed momenta. Overall the convergence scales as $1/n^2$ as opposed to the inverse linear behaviour of standard finite-cluster techniques.  Our results for a square lattice reproduce all known features from state-of-the-art simulations also with only a few mixed momenta. Consequently, we believe the MMHK model offers an alternative tool for strongly correlated quantum matter.
\end{abstract}
\date{\today}

\begin{document}

\flushbottom
\maketitle

\section*{Introduction}

The paradigmatic model for strong correlations in quantum matter is that of electrons hopping on a lattice between nearest neighbors but paying the energy cost $U$ anytime they doubly occupy the same site.  Naively, the energy eigenstates of this model would seem to be ordered with respect to the number of doubly occupied sites.  However, this fails because the kinetic and potential terms do not commute.  As a result, electrons can lower their energy by hopping to sites with single occupancy. It is for this reason that this simple model is unsolved, except in $d=1$\cite{bethe} and $d=\infty$\cite{DMFT}, thereby rendering the d=2 Mott problem a grand challenge as it is the gold standard for Mott physics in the cuprates.  The state-of-the-art methods remain quantum Monte Carlo (QMC)\cite{ZhengScience2017,Huangnpj2018,QinPRX2020,shiwei1,shiwei2}, density matrix renormalization group (DMRG)\cite{Jiangpnas2022,ZhengScience2017,Huangnpj2018,QinPRX2020} and dynamical mean-field theory (DMFT)\cite{DMFT,XuPRL2013,DengPRL2013} and its cluster extensions such as cellular DMFT\cite{ParkPRL2008,KancharlaPRB2008} and the dynamical cluster approximation (DCA)\cite{maier,Maipnas2022,MaiNC2023,WernerPRB2009}, as well as diagrammatic extensions\cite{RohringerRMP2018,LeBlancPRX2015} such as dynamical vertex approximation (D$\Gamma$A) \cite{SCHAFER2016107,ShafferPRX2021} and dual fermion (DF) \cite{RubtsovPRB2008}, which have all been useful in unveiling the properties of the Hubbard model including the pseudogap\cite{maier,kotliar,pg3,pg6,pg9,ShafferPRX2021,RubtsovPRB2008}, superconductivity\cite{QinPRX2020,maier} and transport\cite{HuangScience2019,DengPRL2013,BrownScience2019} in the strange metal regime. 


 A natural question arises:  Is there an alternative that allows a natural interpolation between the two solvable limits of $d=1$\cite{bethe} and $d=\infty$\cite{DMFT}.
 It is this problem that we tackle here. The more tractable formulation of Mott physics is the Hatsugai-Kohmoto (HK) model\cite{HK,HKnp1,HKnp2,Zhao_2025,SkolimowskiPRB2024,MaPRB2025} which we refer to as the band HK model in the following. It has a repulsion between any two electrons with opposite spin that doubly occupy the same momentum state. Consequently, the doubly occupied band steadily moves upwards in energy as the repulsion increases resulting in an insulating state in a half-filled band when the interaction strength exceeds the bandwidth.  In this model, double occupancy is the organizing principle for the ordering of the eigenstates. 
Clearly then, this is a gross simplification of the Hubbard model stemming from the locality in momentum space (long-range interactions in real space) and generating a macroscopic spin degeneracy in the singly occupied sector.  In essence, the solvability of band HK rests on the commutativity of the kinetic and interaction terms.  Can such non-commutativity be put back in systematically, thereby reducing the difference from the Hubbard model, without giving up on exact solvability completely? 

We show here that this can be done by constructing the momentum-mixing Hatsugai-Kohmoto (MMHK) model. The procedure is to group the momenta into cells and then mix them.  For example, grouping the k-states into cells of two mixes all momenta differing by $k=(0,0)$ and $(\pi,\pi)$.  We generalize this to $n$ momenta being mixed by decorating each k-state by $n$ sites, which we denote as the $n$-MMHK model. We show that just a few mixing momenta are needed to recover all known results in 1d\cite{bethe} and is in agreement with standard results in two dimensions (2d) for the Mott transition\cite{DMFT,maier}, spin susceptibility\cite{White}, spectral function\cite{Sekiprb2016,Huangprr2022}, double occupancy\cite{White,shiweizhang}, dynamical spectral weight transfer\cite{sawatzky,Sawatzkyprl}, and heat capacity\cite{Duffyprb1997,Wangprb2022}. We show that in 1d, the convergence to Hubbard physics in the thermodynamic (TD) limit scales as $\sim1/n^2$, whereas in finite-cluster Hubbard calculations, the convergence as a function of system size is inversely linear. The nature of the rapid convergence is explained by a scaling argument in the penultimate section and we show that it becomes even faster in higher dimensions because quantum fluctuations are suppressed. In essence, the MMHK model is an efficient simulator of Mott/Hubbard physics, enabling analytical insights and numerical augmentation.  This is a consequence of the flow of MMHK to the Mott fixed point in which a discrete $Z_2$ symmetry is broken\cite{ppz2}.


\section*{Towards Hubbard}
\begin{figure}[ht]
\centering
\includegraphics[width=0.8\textwidth]{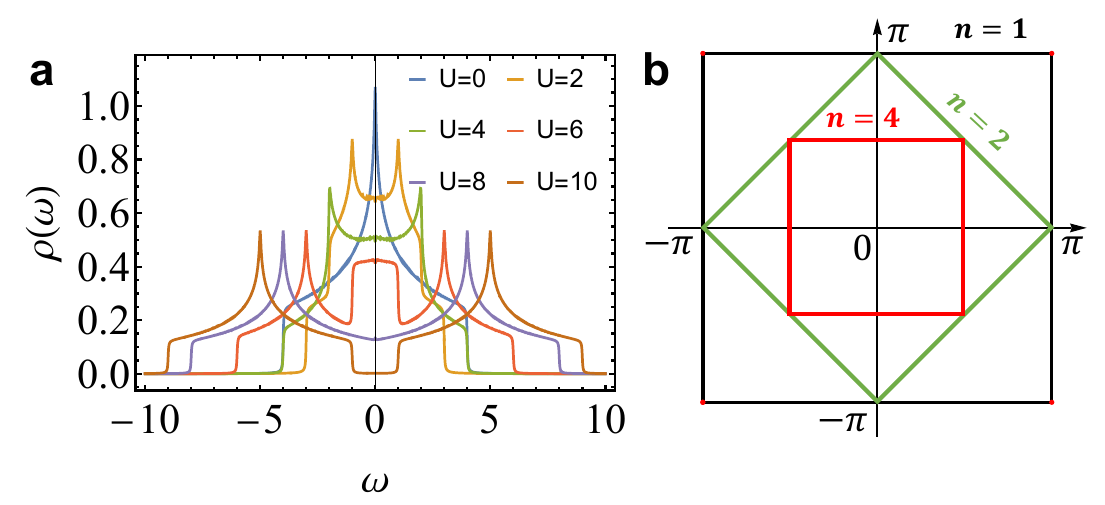}
\caption{ {\bf Mott transition in band HK  model and the scheme for the $n$-MMHK model.} {\bf a}, The DOS under different $U$ displaying the Mott transition in the half-filled band HK model at inverse temperature $\beta=200/t$. {\bf b}, Evolution of the reduced Brillouin zone as the number of mixed momenta $n$ increases, leading to a purely local (in real space) model when $n= N$.}
\label{fig:wilsonhk}
\end{figure}

At the outset, it is helpful to write the explicit form of the Hubbard and HK models.  While both models share the same kinetic energy, $\sum_{\bf k,\sigma}\xi_{\bf k} n_{\mathbf{k}\sigma}$, their potential energy terms,
\begin{equation}
\begin{split}
    H^{\rm HB}_{\rm int}&=U\sum_i n_{i\uparrow}n_{i\downarrow} \\
    &=\frac{U}{N}\sum_{{\bf k},{\bf p},{\bf q}\in \mathrm{BZ}} c^\dagger_{{\bf k}\uparrow} c_{{\bf k}-\bf q\uparrow}c^\dagger_{{\bf k}+{\bf p}\downarrow}c_{{\bf k}+{\bf p}+{\bf q}\downarrow},
\end{split}
\end{equation}

and 
\beq
H^{\rm HK}_{\rm int}=U\sum_{{\bf k}\in \mathrm{BZ}} n_{{\bf k}\uparrow}n_{{\bf k}\downarrow} \label{bHK}
\eeq
differ substantially. In Hubbard, antiparallel spin electrons with local on-site occupancy, $n_{i\sigma}$, repel one another, while in HK doubly occupying the same \emph{momentum} state turns on the interaction.  The occupancy in each momentum state is $n_{\mathbf{k}\sigma}$ and $\xi_{\mathbf{k}}$ is the energy of each band state with momentum $k$. It is easy to see that because the HK model is diagonal in momentum space, should $U$ exceed the bandwidth set by $\xi_{\mathbf{k}}$, a gap opens at the chemical potential and the spectral weight will now be carried by two bands each with equal weight at half-filling.  It is this bifurcation of the spectral weight that is central to Mott physics as depicted in Fig. (1) of \refdisp{ppfixedp}. \figdisp{fig:wilsonhk}(a) demonstrates that the density of states (DOS) for various $U$ across the Mott transition of the half-filled band HK model always exhibits a finite DOS at $\omega=0$ for $U<W$.  This central peak only vanishes in the insulating phase.   In HK, there are 4 states per k-point.  None of the k-points are hybridized.  To bridge the gap with Hubbard, we must introduce hybridization.  The simplest way to bridge this gap is to put two k-points into a unit cell, and then hybridize them. The difference between the hybridized momenta will be  $B_2\equiv \{(0,0),(\pi,\pi)\}$.  In this basis, the effective interaction can be written in the standard Hubbard form 
\beq
    H^{\rm MMHK}_{{\rm int}, n=2}=\frac{U}{2}\sum_{\mathbf{k}\in \mathrm{BZ}}\sum_{\mathbf p\in B_2}\sum_{\mathbf q\in B_2}c_{\mathbf{k}\uparrow}^\dagger c_{\mathbf{k}-\mathbf q\uparrow}c_{\mathbf{k+p}\downarrow}^\dagger c_{\mathbf{k+p+q}\downarrow}. \label{mm}
\eeq
However, by bipartitioning the lattice and introducing the basis for each unit cell (sums and differences of the k-state fermionic operators), $c_{\mathbf{k}A\sigma}=\frac{1}{\sqrt{2}}\left(c_{\mathbf{k}\sigma}+c_{\mathbf{k}+(\pi,\pi)\sigma}\right)$ and $c_{\mathbf{k}B\sigma}=\frac{1}{\sqrt{2}}\left(c_{\mathbf{k}\sigma}-c_{\mathbf{k}+(\pi,\pi)\sigma}\right)$, we find that this potential can be written in a compact form,
\beq
    H^{\rm MMHK}_{{\rm int}, n=2}=U\sum_{\mathbf{k}\in \mathrm{rBZ}_2}\left(n_{\mathbf{k}A\uparrow}n_{\mathbf{k}A\downarrow}+n_{\mathbf{k}B\uparrow}n_{\mathbf{k}B\downarrow}\right),
\eeq
reminiscent of the HK interaction \disp{bHK} but with $\mathbf{k}$ and $\mathbf{k}+(\pi,\pi)$ mixed ($n=2$). In doing so, we must restrict the momentum summation to the reduced Brillouin zone, rBZ$_2$ which is exactly half the original Brillouin zone (BZ).  Then for $n=2$, we hybridize the k-points that differ by $(\pi,\pi)$, thereby lifting their degeneracy.  This procedure, which can be repeated with any size of unit cell, continuously deforms HK into the Hubbard model.  The question we ask is can this procedure be automated and how large does the unit cell have to be to reproduce non-trivial Hubbard physics?

\section*{MMHK} 

We now outline how the k-point hybridization picture can be automated.
As pointed out previously\cite{barry}, band HK can be generalized to include multiple orbitals per ${\bf k}$-state as is necessary for the adaptation of this model to topological models, all of which have at least 2 atoms per unit cell\cite{pphaldane,ppqnh,KrystianPRB2024}.  Such a change produces qualitatively different physics as the hybridization between the orbitals lifts\cite{barry,JablonowskiPRB2023,ZhongPRB2022} the thermodynamic degeneracy of the band HK model while preserving only the degeneracy constrained by the symmetry\cite{Settysym}.  The procedure we outline here is equivalent to this scheme only if the number of degrees of freedom is fixed as orbitals are added.  That is, the BZ with $n$ orbitals per site is reduced by a factor of $1/n$.  Since there is no constraint of this kind inherent in the orbital model\cite{barry}, we refer to the scheme adopted here as the MMHK model. As we will see, the convergence of MMHK to the Hubbard physics in the TD limit scales as fast as $\sim1/n^2$.

Since we are only interested in Hubbard physics for now, we consider an MMHK model of the form,
 \begin{equation}
     \begin{split}
         H^{\text{MMHK}}_n=& \sum_{{\bf k},\alpha,\alpha',\sigma}g_{\alpha\alpha'}({\bf k})c_{{\bf k}\alpha\sigma}^\dagger c_{{\bf k}\alpha'\sigma}-\mu\sum_{{\bf k},\alpha}\hat{n}_{{\bf k}\alpha\sigma} \\&+\sum_{{\bf k},\alpha} U \hat{n}_{{\bf k}\alpha\uparrow} \hat{n}_{{\bf k}\alpha\downarrow},
 \label{ohk}
     \end{split}
 \end{equation}
where $n$ is the number of mixed momenta, $k$ is summed over the rBZ$_n$ and $g(\bf k)$ is the dispersion matrix whose elements in a tight-binding model are defined by $g_{\alpha\alpha'}=\sum_{\mathbf{\delta r}}t^{\alpha,\alpha'}_{\mathbf{\delta r}}e^{i\mathbf{k}\cdot(\mathbf{r_\alpha-r_{\alpha'}+\delta r})}$ where $\mathbf{r}$ denotes the position of the unit cell and $t^{\alpha,\alpha'}_{\mathbf{\delta r}}$ represents the hopping between sites $\alpha$ and $\alpha'$ located in unit cells separated by $\mathbf{\delta r}$ and $\sigma$ is the spin. 
While the kinetic and interaction terms no longer commute, the Hamiltonian still maintains the form $\sum_{\mathbf{k}\in \rm{rBZ_n}} h_{\mathbf{k}}$, is governed by the HK fixed point\cite{ppfixedp,ppz2} and still remains exactly solvable when the unit cell is not too big. Since the Hamiltonian in \disp{ohk} depends explicitly on $n$, we also denote it as the $n$-MMHK model. Of course, the solvability is now based on finding the eigenstates of an $n$-site cluster. \figdisp{fig:wilsonhk}(b) depicts the reduction of rBZ when $n$ sites decorate each k-point (see Extended Data Fig.~1 for real-space counterparts). In the limit of $n\rightarrow N $, the k-summation vanishes reducing the interaction term entirely to,  
\beq
\begin{aligned}
\label{ninfty}
&\lim_{n\rightarrow  N }\sum^n_{\alpha=1} \sum_{{\bf k}\in {\rm rBZ}_n} U n_{{\bf k}\alpha\uparrow} n_{{\bf k}\alpha\downarrow} =\sum^{ N }_{\alpha=1} U n_{\alpha\uparrow} n_{\alpha\downarrow},
\end{aligned}
\eeq
which is just a summation over the local degrees of freedom as in the on-site Hubbard model. Hence, we are guaranteed to obtain Hubbard physics for $n$ sufficiently large. 

\section*{One dimension}

\begin{figure}[bt!]
    \centering
    \includegraphics[width=0.9\textwidth]{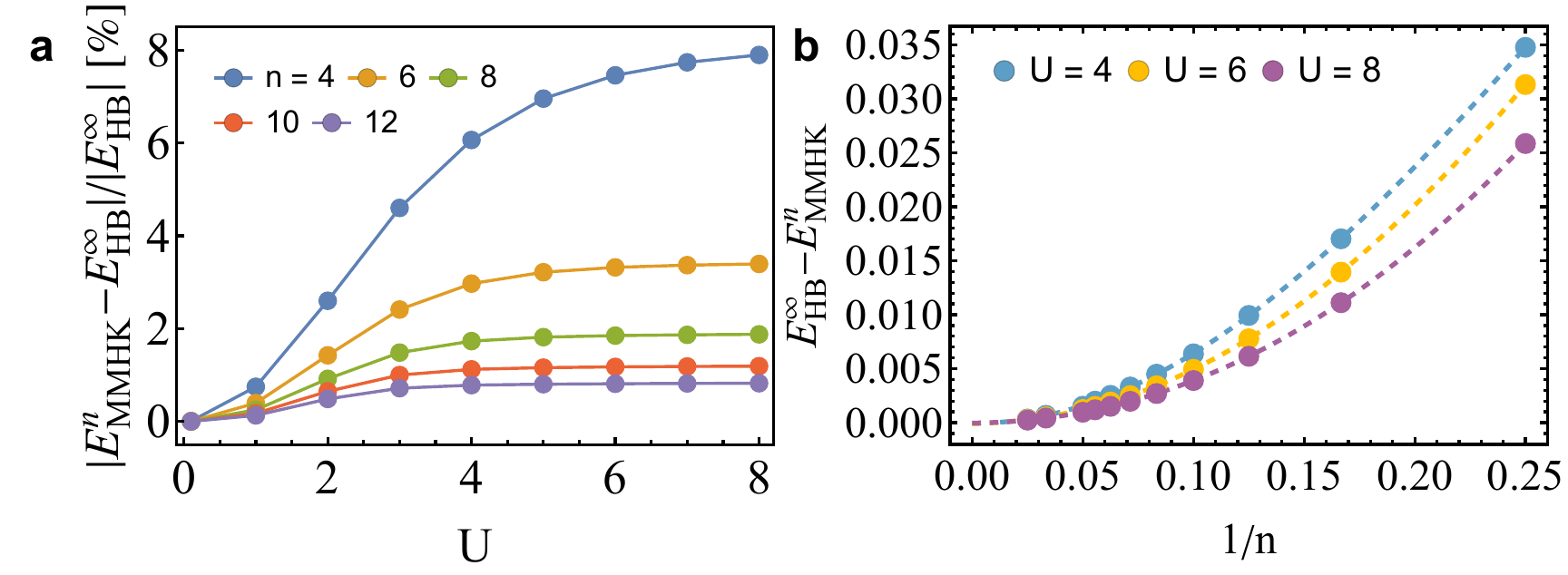}
    \caption{{\bf Ground-state energy deviation of $n$-MMHK from Bethe ansatz.} Comparison of the difference between the ground-state energy with $n$-MMHK model, $E_{\rm HK}^n$, and the infinite-system size Hubbard Bethe ansatz energy, $E_{\rm HB}^\infty$, both in one dimension. {\bf a}, MMHK is solved by exact diagonalization. {\bf b}, $n$-MMHK is solved by DMRG including as many as $n=40$ mixed momenta at various $U$. The dashed lines are polynomial regression fitting with extrapolation to $1/n\rightarrow 0$ ($n\rightarrow\infty$). The fitting curves can be well represented as $f(U=4)=0.45(1/n^{1.83})+a$, $f(U=6)=0.51(1/n^{2.01})+a$ and $f(U=8)=0.45(1/n^{2.07})+a$. The asymtotic values $a$ at $1/n=0$ are $-5.4\times10^{-5}~(U=4)$, $-9\times10^{-5}~(U=6)$, $-4\times10^{-5}~(U=8)$. 
    }
    \label{exact}
\end{figure}

\figdisp{exact}a presents a comparison of the ground-state energy vs $U$ between the exact Bethe ansatz result of the one-dimensional (1D) Hubbard model and that of $n$-MMHK with varying $n$). As $n$ increases, the $n$-MMHK model rapidly approaches the exact result\cite{bethe}.  Already at $n=10$, the deviation from the exact result is less than $1\%$. Note that even for relatively small $n$, it converges to the exact Hubbard results consistently as $n$ increases across all values of $U$. This behavior stands in sharp contrast to finite-Hubbard-chain calculations, where deviations from the exact results vary unpredictably with both chain length $L$ and $U$ when $L$ is small (Extended Data Fig.~2). This suggests that a small $n$ can capture the essential Hubbard physics qualitatively, with increasing $n$ offering a systematic path to quantitative accuracy. 

To further quantify the rate of convergence with increasing $n$, we solve the $n$-MMHK model using DMRG with $n$ as large as 40. In \figdisp{exact}b, we plot the difference between the $n$-MMHK ground-state energy and the exact Bethe ansatz result for the Hubbard model in the TD limit as a function of $1/n$ for $U/t=4, 6, 8$. Rather than scaling as $1/n$, we find that the actual convergence follows a power law $1/n^\gamma$ with $\gamma\approx2$. Specifically, $\gamma=1.83$ for $U=4$, increasing with $U$ and reaching $\gamma=2.07$ for $U=8$. In contrast, the standard finite-size Hubbard chain calculation using DMRG with open boundary condition (OBC) only shows a scaling of $1/L$ (Extended Data Fig.~3). These results indicate that the $n$-MMHK model offers a more rapid convergence toward Hubbard physics in the TD limit than do finite-cluster Hubbard calculations. Notably, we later observe an even faster convergence rate than $1/n^2$ when extending the $n$-MMHK framework to a quasi-2d ladder geometry (Extended Data Fig.~4). This consistent convergence behavior enables data-driven techniques to efficiently extrapolate to the $n\rightarrow  N$ limit. We also compare the single-particle charge or Mott gap, $\Delta$, between both $n$-MMHK-DMRG and Hubbard-DMRG (Extended Data Fig.~5). In $n$-MMHK, $\Delta$ vanishes when $U\rightarrow 0$, as expected, whereas it remains finite in Hubbard-DMRG as a result of finite size effects. This limitation arises because OBC, regardless of the system size, cannot sample the $k=0$ point which is the momentum at which the gap is minimized in $d=1$.  In contrast, $n$-MMHK is constructed to directly represent the TD limit, and as a result, it correctly captures the vanishing of the gap as $U\rightarrow0$. Finally, the $n$-MMHK-DMRG calculations involve a smaller number of mixed momenta compared to standard Hubbard-DMRG chain for the same accuracy, due to the intrinsic periodic boundary conditions imposed by the $n$-MMHK Hamiltonian. Since DMRG performs worse with PBC (typically requiring about five times the bond dimension compared to OBC), this requirement limits the accessible number of mixed momenta. Nonetheless, our 1D study provides a clear and controlled demonstration of the $n$-MMHK model’s rapid convergence. In the following sections, we turn to the real challenge: applying $n$-MMHK to two-dimensional systems.

\section*{Two dimensions}

Applying the $n$-MMHK model to 2D Hubbard simply requires arranging the $n$ sites in a cluster. The consistent and rapid convergence observed in 1D chain and quasi-2d laddder give us reason to expect that even a small number of sites can capture essential Hubbard physics, independent of dimensionality. In the following, we confirm this idea by applying the $4$-MMHK model (where ``$4$" denotes the number of mixed momenta) to 2D, and demonstrate that further quantitative accuracy in the spectral function is achieved with the $16$-MMHK model (with $4\times4$ cluster for each {\bf k}). For $n=4$, implemented here as a $2\times 2$ cluster, the matrix $g({\bf k})$ in \disp{ohk} is
 \beq
 \left(\begin{array}{cccc}
 0 & \varepsilon_{tx} & \varepsilon_{ty} & \varepsilon_{t^\prime}\\
 \varepsilon_{tx} & 0 & \varepsilon_{t^\prime} & \varepsilon_{ty}\\
 \varepsilon_{ty} & \varepsilon_{t^\prime} & 0 & \varepsilon_{tx} \\
 \varepsilon_{t^\prime} & \varepsilon_{ty} & \varepsilon_{tx} & 0
 \end{array}\right),\label{2x2}
 \eeq
with $\varepsilon_{tx(y)}=-2t\cos k_{x(y)}$ and $\varepsilon_{t^\prime}=-4t^\prime \cos k_x\cos k_y$, where $t$ and $t'$ represent the first- and second-neightbor hopping, respectively. The matrix $g({\bf k})$ for $3\times 3$ cluster ($9$-MMHK) is given in the supplementary Eq.~(S1) and Fig.~S1 and straightforward to generalize to any cluster. We set $t'=0$ unless specified. This matrix is identical in form to that of a $2\times 2$ Hubbard cluster with twisted boundary conditions (TBC) \cite{ehpg}, where $\alpha$ is replaced with the site index $i$ and $k_x=\theta_x$ and $k_y=\theta_y$ where $\theta_x$ and $\theta_y$ define the TBC.  With $n$-MMHK, however, the twist has a physical interpretation as the crystal momentum, thereby placing $n$-MMHK in the thermodynamic limit. 

\begin{figure}[t!]
\centering
\includegraphics[width=0.8\textwidth]{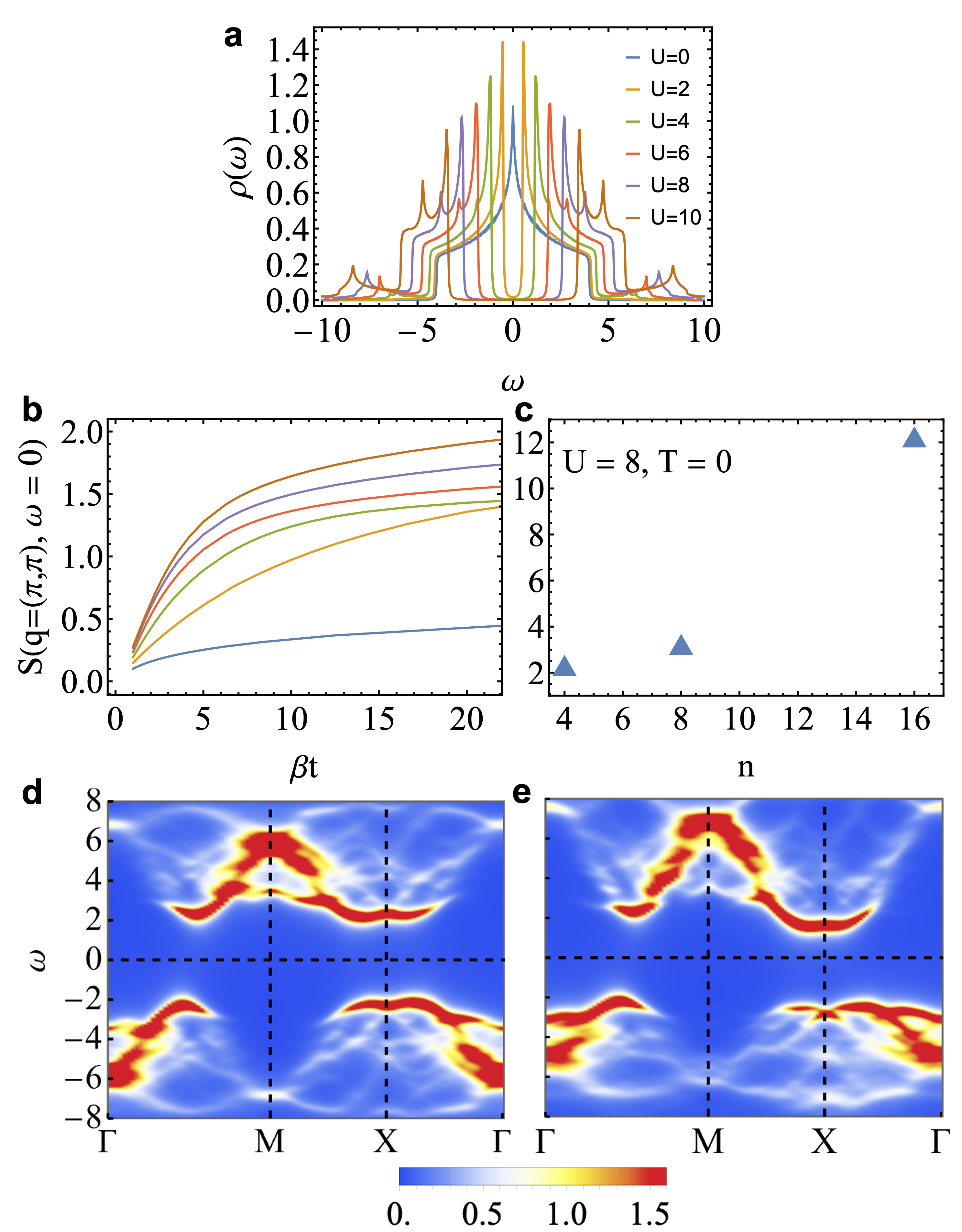}
\caption{{\bf Mott transition, antiferromagnetic susceptibility and spectral functions for the half-filled $n$-MMHK model.} {\bf a}, DOS representing the Mott transition for the half-filled $4$-MMHK model at $\beta=200/t$. {\bf b}, Antiferromagnetic spin susceptibility of the half-filled $4$-MMHK model plotted as a function $\beta$ for various $U$ (same legend as in {\bf a}). Its zero-temperature limit is shown in panel {\bf c} at $U=8$ for $4$-, $8$- and $16$-MMHK. {\bf d,e} show Spectral function $A({\rm k},\omega)$ at half-filling of the $16$-MMHK model at $t'=0$ and $t'/t=-0.25$ respectively, with $U=8$, zero temperature, and a broadening factor of $0.2$.}
\label{fig:MottT}
\end{figure}

We compute the DOS of half-filled $4$-MMHK for a square lattice at $t'=0$ and find that any non-zero $U$ is sufficient to eliminate the metallic state producing a vanishing DOS at zero energy (\figdisp{fig:MottT}a). This reflects the instability caused by Fermi surface nesting. This is captured already in $4$-MMHK--a qualitative improvement from band HK in which $U_c=W$ (bandwidth)--and is missed by DMFT\cite{DMFT}, though cluster DMFT\cite{ParkPRL2008} progressively yields a smaller $U_c$ as the cluster size increases, consistent with results from DCA\cite{MaierRMP2005} and D$\Gamma$A\cite{SCHAFER2016107}. For finite $t'$, which breaks nesting, a finite $U_c$ is needed to open a charge gap (Extended Data Fig.~6). Another significant improvement over the band HK model, which exhibits a diverging ferromagnetic susceptibility, is that the $4$-MMHK model shows antiferromagnetic (AF) correlations as the leading spin response (Extended Data Fig.~7), a key feature of Hubbard physics. In Fig.~\ref{fig:MottT}b, we show that the AF spin susceptibility $S(q=(\pi,\pi),\omega=0)=\int_0^\beta d\tau\langle \hat{S}^z(q,\tau)\hat{S}^z(0)\rangle$ increases with 
$U$ and decreasing temperature. The Mott insulating state can be reached either by lowering the temperature at large $U$ or by increasing $U$ at low temperature. In two dimensions, AF ordering, prohibited at finite temperature by the Mermin-Wagner theorem\cite{mermin}, is not crucial to the temperature-driven transition at energy scales $\sim U$, well above the AF exchange scale of $t^2/U$. In contrast, at low temperature with finite $t'$, the AF susceptibility shows a cusp at $U_c$ (Extended Data Fig.~6), indicating that AF fluctuations are present in the interaction-driven Mott transition. The persistence of AF correlations away from half-filling (Extended Data Fig.~7) further highlights their key role in doped Hubbard physics.
The AF spin susceptibility (\figdisp{fig:MottT}b) is underestimated in the $4$-MMHK model--where the system consists of Hubbard squares in $k$ space--compared to
 QMC results on larger Hubbard clusters \cite{ShafferPRX2021,QinAnnrev2022}, and it does not diverge at zero temperature. This quantitative discrepancy can be mitigated by including additional mixed momenta (\figdisp{fig:MottT}c). 
We then compute the spectral function $A({\rm k},\omega)$ at $U/t=8$ of the half-filled $16$-MMHK (with $4\times4$ cluster per ${\bf k}$) using Lanczos exact diagonalization (ED) for $t'/t=0$ and $-0.25$, shown in \figdisp{fig:MottT}(d) and (e) respectively. The narrowing dispersion of the lower and upper Hubbard bands are in quantitative agreement with state-of-the-art QMC and cluster pertubation theory for the 2d Hubbard model\cite{Sekiprb2016,Huangprr2022,Schummarxiv2025} at $t'=0$ (see Extended Data Fig.~8 for the $4$- and $8$-MMHK results which already captures the qualititive features). For $t'=-0.25$, unbiased QMC suffers from the sign problem at low temperatures. Our result (\figdisp{fig:MottT}(e)) exhibits particle-hole asymmetry and an indirect Mott gap. This indicates that doped holes preferably occupy regions around $(\pi/2,\pi/2)$, while doped electrons appear first near $(0,\pi)$ and $(\pi,0)$, laying a solid foundation for future studies on particle-hole asymmetric pseudogap physics. 

\begin{figure}[t!]
	\centering\includegraphics[width=\textwidth]{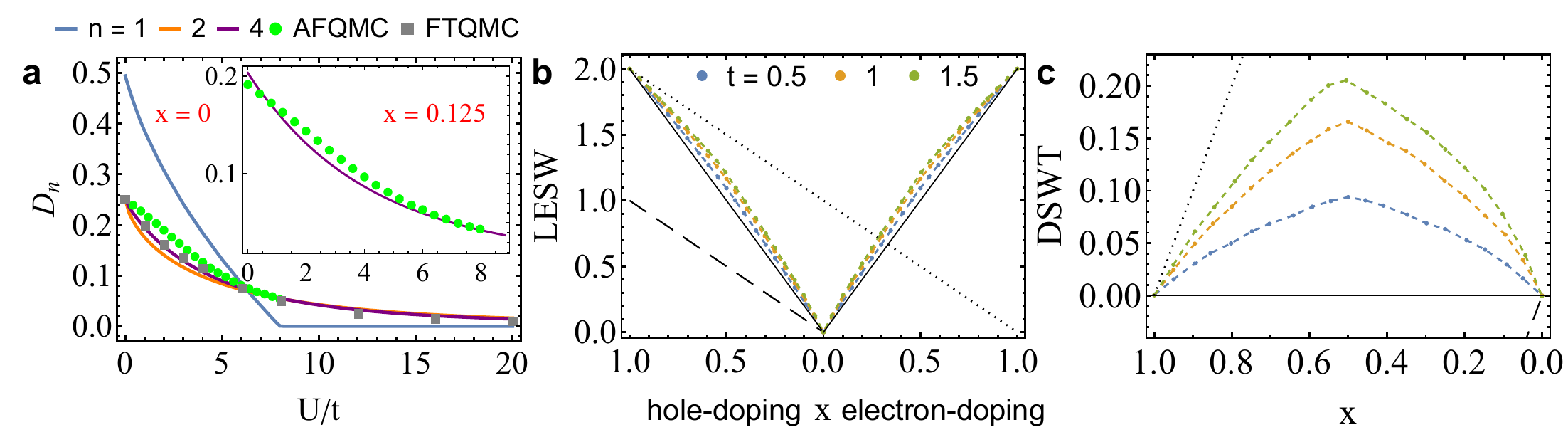}
	\caption{{\bf Double occupancy and dynamical spectral weight transfer for $n$-MMHK.} {\bf a}, $D_n$ at half-filling ($x=0$) as a function of $U$ for the $n$-MMHK model with various $n$. The green and gray dots are from auxiliary-field QMC (AFQMC)\cite{shiweizhang} and finite-temperature QMC (FTQMC) calculations\cite{White} of the standard double occupancy of the 2D Hubbard model. The inset of {\bf a} compares the $4$-MMHK and Hubbard-AFQMC results\cite{shiweizhang} at $1/8$-hole-doping ($x=0.125$). LESW and DSWT are the exact solution of the $4$-MMHK model ($U=10$) for the hopping parameters shown in {\bf b} and {\bf c} respectively at $\beta=30$. The solid line shown with slope $2x$ is the band HK or atomic Hubbard result.  The dashed and dotted lines depict the semiconductor and Fermi liquid results respectively. Note, there is no qualitative difference with the ED results for the Hubbard model\cite{sawatzky,Sawatzkyprl}. As $t/U$ increases, so does the DSWT.}
	\label{dynamicsw}
\end{figure}

Given that the kinetic and potential terms do not commute, the number of doubly occupied sites plays a crucial role in all the ground state properties.
To this end, we compute the double occupancy for the $n$-MMHK defined as
 \beq
 D_n = \frac{1}{N} \sum_{{\bf k}\in \rm{rBZ_n},\alpha} \langle n_{{\bf k}\alpha\uparrow} n_{{\bf k}\alpha\downarrow} \rangle,
 \label{eq:doccu}
 \eeq
 which is directly related to the interaction energy. Consider the stark difference between the non-interacting limits of the band HK and $n$-MMHK models in terms of double occupancy.  Regardless of the model, in the non-interacting limit, the expression for $D_n$ factorizes 
 \beq
 D^{\rm non-int}_n=\frac{1}{N} \sum_{{\bf k}\in \rm{rBZ_n},\alpha} \langle n_{{\bf k}\alpha\uparrow}\rangle\langle n_{{\bf k}\alpha\downarrow} \rangle.
 \eeq
For the band HK model ($n=1$), half the BZ is doubly occupied.  That is, $\langle n_{{\bf k}\alpha\sigma}\rangle=1$ only if $k\alpha$ is occupied.  Otherwise, $\langle n_{{\bf k}\alpha\sigma}\rangle=0$.  Consequently, $D^{\rm non-int}_1=1/2$.  On the other hand, for the $n$-MMHK model, $\langle n_{{\bf k}\alpha\sigma}\rangle=\langle n_{{\bf k}\beta\sigma}\rangle$ for all $\alpha\neq\beta$ by rotational symmetry. In addition, $\langle n_{{\bf k}\alpha\sigma}\rangle=1/2$ in the full range of the rBZ. As a result, we find that for the $n$-MMHK model, $D^{\rm non-int}_{n>1}=1/4$. Note that this result is identical to the non-interacting limit of the Hubbard model and underscores how drastically band HK differs from its momentum-mixing counterpart. \figdisp{dynamicsw}a for $D_n$ shows this dramatic difference.  In band HK, $D_1$ decreases steadily from $1/2$ and vanishes for $U>W$.  However, in $n$-MMHK model, $D_n$, starting at $1/4$, just tapers asymptotically as $U$ increases.  Note also the rapid convergence between the $n=2$ and $4$ cases which match closely with state-of-the-art auxiliary-field QMC simulations (AFQMC)\cite{shiweizhang} and finite-temperature QMC (FTQMC) \cite{White} on the Hubbard model. The inset also shows the double occupancy of the $4$-MMHK model at $1/8$-hole-doping ($x=0.125$) and its benchmark with QMC simulations\cite{shiweizhang} (see supplementary information for explanation of the slight mismatch at $U=0$ due to definition).  This agreement here underscores that $n$-MMHK converges rapidly to Hubbard physics with negligible computation effort. 
 

 The fact that the double occupancy only vanishes asymptotically as $U$ increases in the $n$-MMHK model reflects the dynamical mixing between the upper and lower Mott sub-bands.  This arises entirely from the non-commutativity\cite{harrislange,sawatzky,Sawatzkyprl} of the kinetic and potential energies and appears as $t/U$ corrections to the low-energy spectral weight (LESW).  In the atomic limit, the LESW is strictly $2x$ (where $x$ is the doping level) because each hole can be occupied by a spin-up or spin-down electron.  Any dynamical corrections to this necessarily generate double occupancy and hence increase the LESW strictly defined as 
 \beq
 \Lambda(x)\equiv{\rm LESW}=\int _0^{\omega_g} N(\omega)d\omega.
 \eeq  
 The $\omega_g$ locates at the spectral gap in the DOS (Extended Data Fig.~9). In \figdisp{dynamicsw}b, we compare the LESW of the $4$-MMHK model with Fig.~3 of \refdisp{sawatzky}. There is no qualitative difference with ED on the (1D) Hubbard model\cite{sawatzky,Sawatzkyprl}. Both increase faster than $2x$.  The semiconductor (dashed line) and Fermi liquid (dotted line) results are shown for comparison.  As the occupied part of the lower band has a weight $1-x$, the total weight in the lower band now exceeds $1+x$.  As only $1+x$ electrons can occupy the lower band, dynamical spectral weight transfer (DSWT), defined as $\Lambda(x)-2x$ and plotted in \figdisp{dynamicsw}c, implies\cite{rmpphillips} that the spectral weight in the lower band cannot be exhausted by counting electrons alone. \figdisp{dynamicsw}c is quantitatively in agreement with the Hubbard model (Fig.~4 of \refdisp{sawatzky}), the only difference being the maximum value which for $4$-MMHK is $0.2$ whereas it is $0.23$ for the (1D) Hubbard result. Consequently, that the LESW in the $n$-MMHK model increases faster than $2x$ is a profoundly non-trivial result as no analytically solvable model has ever captured this feature of the Hubbard model. The ability of the $n$-MMHK model to reproduce this behavior with as few as $n = 2$ or $4$ mixed momenta offers a rare opportunity to gain analytical insight into DSWT, as demonstrated in a follow-up study \cite{TenkilaPRB2025}.

\begin{figure}[t!]
	\centering
	\includegraphics[width=0.75\textwidth]{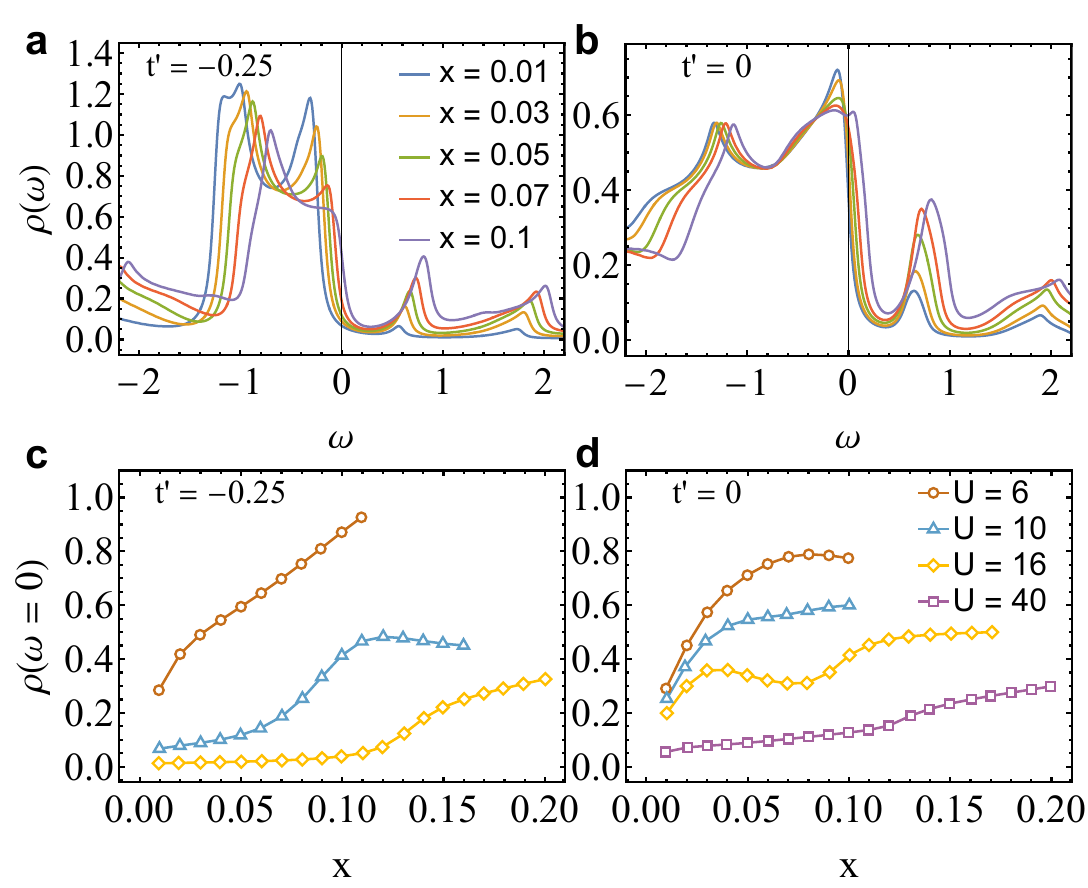}
	\caption{{\bf Observation of pseudogap in $4$-MMHK.} DOS at varying hole-doped densities with ({\bf a}) and without ({\bf b}) the next-nearest-neighbor hopping $t'$ for the $4$-MMHK model ($U/t=10, \beta=30/t$). {\bf a} and {\bf b} share the same legend. Only for $t'\ne 0$ is $\rho(\omega)$ suppressed at zero frequency in the under-doped region, thereby indicating a pseudogap. {\bf c} and {\bf d} show $\rho(\omega=0)$ with $t'=-0.25$ and $0$ respectively, as a function of hole-doped density under various $U$.}
	\label{fig:pgp}
\end{figure}

Aside from DSWT, $n$-MMHK also exhibits a pseudogap in the DOS.  Shown in \figdisp{fig:pgp} is the DOS for the $4$-MMHK model with $t'=-0.25$ (\figdisp{fig:pgp}(a,c)) and $t'=0$ (\figdisp{fig:pgp}(b,d)). \figdisp{fig:pgp}(a) shows a suppression of the DOS at the chemical potential ($\rho(\omega=0)\lesssim 0.1$) in the underdoped region with $t'=-0.25$ when $U\geq W$, indicative of a pseudogap in the absence of superconducting order. The complete tracking of the DOS at zero frequency in \figdisp{fig:pgp}(c) displays the trend that as $U$ increases, the pseudogap region appears and extends to higher hole-doped density. In contrast, when $t'=0$, the pseudogap suppression does not arise until $U\gg W$, as shown in \figdisp{fig:pgp}(b,d). These observations are consistent with a recent study\cite{ehpg} on a $2\times 2$ Hubbard cluster with TBC. The emergence of the pseudogap upon including these scattering momenta highlights their crucial role in suppressing spectral weight, consistent with a recent solvable model construction \cite{WormPRL2024}.

\begin{figure}[t!]
	\centering\includegraphics[width=0.75\textwidth]{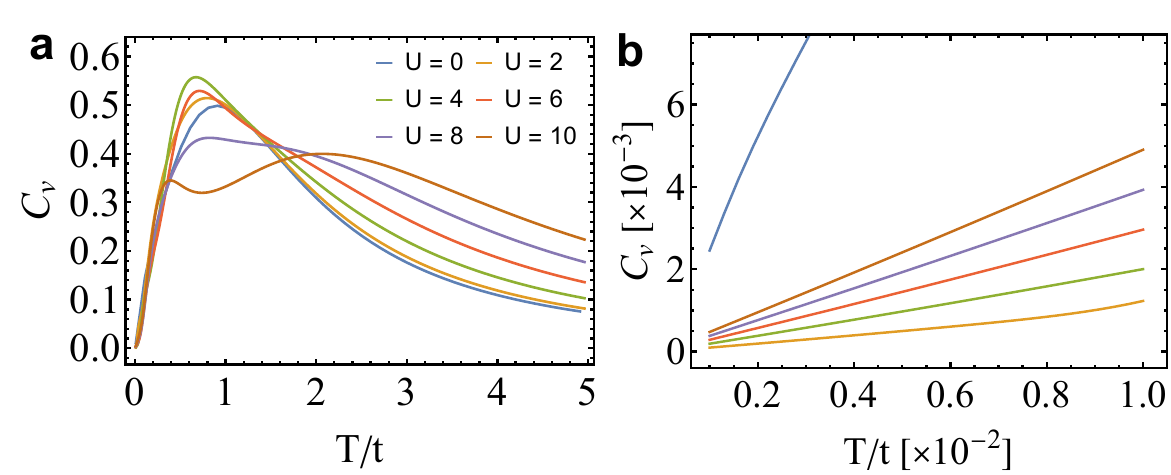}
	\caption{{\bf Heat capacity of the half-filled $4$-MMHK model.} {\bf a, b} present the heat capacity as a function of temperature at high and low temperatures respectively for various $U$. {\bf a, b} share the same legend.}
	\label{HC}
\end{figure}

Finally, we compute the heat capacity of the $4$-MMHK model at half-filling shown in \figdisp{HC} Additional results for the $1$‑ and $2$‑MMHK cases are provided in Extended Data Fig.~10. Most noticeable is the two-peak structure at $U>W$, representing a demarcation of the charge and spin degrees of freedom into high and low-temperature regimes, respectively, consistent with QMC simulations on the Hubbard model \cite{Duffyprb1997,Wangprb2022}.  Also at half-filling, we find a near-crossing of the heat capacity curves as a function of $U$ at a temperature intermediate between the spin and charge excitations.  The Maxwell relations governing the entropy dictate\cite{vollhardt} the existence of such a crossing.   Since there is no sign-problem restriction, we can access the low-temperature heat capacity exactly.  As we show in \figdisp{HC}(b), the low-temperature heat capacity data follow a power-law increase detailed in the supplement.  While such algebraic growth might seem counterintuitive for a Mott insulator, the first-excited state is charge-neutral\cite{gnc,fn3} as shown in the supplement. Such charge-neutral excitations determine the low-T behavior of the specific heat and are consistent with the absence of long-range magnetic order.    


\section*{Scaling}

The hidden power of the $n$-MMHK model lies in its rapid and consistent convergence toward Hubbard physics. This efficiency arises from the number of degrees of freedom introduced by the site number at each $k-$point: $n$ sites generate $n$ hybridized momenta, leading to $n^2$ interaction terms at each k-point (\disp{mm}), and hence the scaling $1/n^2$, independent of dimensionality. Including momentum mixing in this way effectively introduces quantum fluctuations. We demonstrate the $1/n^2$ scaling in 1D (\figdisp{exact}), where quantum fluctuations are strongest. In Extended Data Fig.~4, we find even faster convergence $(\sim1/n^3)$ on a quasi-2d ladder system, particularly at larger values of $U$, as a result of suppressed fluctuations in higher dimensions.  This makes two dimensions an ideal setting for the MMHK framework: a computable number of sites can cover a square or rectangular unit cell with sizable linear extent $l$, while quantum fluctuations remain moderate. Furthermore, this argument points towards a monotonically increasing  exponent, $\gamma(d)$ as a function of the lattice  dimension giving rise to a convergence of $1/n^{\gamma(d)}$ where $\gamma(d=1)=2$ and increases for $d>1$. Therefore, vanishingly few mixed momenta are required in $d=\infty$ to achieve convergence, implying a consilience between band HK model and Hubbard model in this limit. As the underlying Mott transition is governed by the breaking of the $Z_2$ symmetry\cite{ppz2}, thereby determining the universality class of the second-order end point, this rapid convergence in $d=\infty$ is highly suggestive of the similarity in criticality between HK and DMFT\cite{DMFT}. 
Since this involves an in depth comparison of the critical exponents, we postpone this for a future study.  

Why does this scheme work so well?  We have previously shown\cite{ppz2} using the Bott topological invariant that the surface of zeros or the Luttinger surface, the defining feature that leads to the charge gap in a Mott insulator, is stable against any continuous local deformation. Since the momentum-mixing construction is based on a $n$-site extension, thereby a local deformation, the Luttinger surface must exist for any $n$. Consequently, as far as the charge dynamics are concerned, HK and Hubbard lie in the same universality class. This perspective is valid because they are both governed by the Mott fixed point of $\mathbb Z_2$ symmetry breaking \cite{ppz2}. The spectral weight bifurcation into upper and lower Hubbard bands is preserved under momentum-mixing extension \cite{ppfixedp}, allowing $n$-MMHK to faithfully capture the essential charge physics of the Hubbard model while incorporating non-trivial dynamics in a controlled and systematic way.

\section*{Concluding Remarks:}  Ultimately, the deep connection between $n$-MMHK and Hubbard physics stems from the resilience of HK to local continuous deformations.  The momentum-mixing construction represents such a local deformation.  Since these deformations leave\cite{ppz2} the Luttinger surface intact, the same universality class governs the charge dynamics that mitigate the Mott transition in both $n$-MMHK and Hubbard models. This applies only to the charge sector; differences arise in the spin sector.  For example, while band HK exhibits a ferromagnetic instability \cite{pphaldane}, the $n$-MMHK and Hubbard models both display antiferromagnetism. This is expected as the $\mathbb Z_2$ breaking fixed point only controls the charge dynamics. After all, it is the charge gap that is the universal feature of all Mott insulators.  This is consistent with Mott's original idea\cite{mott1949} that the spin sector is ancillary to the insulating ground state in partially filled bands. In fact, our fixed point argument\cite{ppfixedp} is a proof of Mott's idea and $n$-MMHK is the numerical demonstration of its validity.

We have shown that the $n$-MMHK model captures key $k$-summed quantities such as the DOS, reflecting the underlying Mott fixed point. This suggests that even with a small number of mixed momenta, $n$-MMHK can potentially provide meaningful insight into broader phenomena like the strange metal phase. For $k$-resolved properties, $n=16$  yields quantitatively accurate spectral functions, enabling detailed studies of Fermi surface reconstruction and the pseudogap. At this scale, spin fluctuations relevant to superconductivity and stripe order are also captured quantitatively.  Additionaly, $n$-MMHK can be used to study transport provided the order of limits is correctly\cite{MaPRB2025} taken, 1) $q\rightarrow 0$ then 2) $L\rightarrow \infty$ rather than the inverse which will yields erroneous\cite{GuerciPRB2025} results. In summary, the rapid and systematic convergence of $n$-MMHK to Hubbard physics makes it a powerful and efficient framework for exploring strong correlations. With just a few mixed momenta, it offers a simplified yet faithful platform to study the interplay between correlations and other effects such as topology and nonequilibrium dynamics. 


\section*{Acknowledgements} 
We thank Barry Bradlyn for helpful comments and Shiwei Zhang for sending us his double occupancy data that is plotted in \figdisp{dynamicsw}a (green dots) and George Sawatzky for a helpful e-mail exchange and Zhentao Wang for his help in exact diagonalization. This work was supported by the Center for Quantum Sensing and Quantum Materials, a DOE Energy Frontier Research Center, grant DE-SC0021238 (P. M. and P. W. P.). PWP also acknowledges NSF DMR-2111379 for partial funding of the HK work which led to these results. P.M. was also supported by the Gordon and Betty Moore Foundation’s EPiQS Initiative through grant GBMF 8691. The DQMC calculation of this work used the Advanced Cyberinfrastructure Coordination Ecosystem: Services \& Support (ACCESS) Expanse supercomputer through the research allocation TG-PHY220042, which is supported by National Science Foundation grant number ACI-1548562\cite{xsede}.

\section*{Author Contributions}
P.M. performed the calculations for $4$- and $16$-MMHK models in two dimensions, and analyzed the data; J.Z. performed the calculations for $2$-MMHK model and carried out the analytic derivation; G.T. performed the DMRG calculations for MMHK in one dimension and ladder; N.A.H. performed the ED calculations for MMHK in one dimension; D.K. performed the calculations for local correlations in the band HK model; D.P. performed the calculations for spin correlations in the band HK model;P.W.P. supervised the project; P.M., J.Z., G.T., and P.W.P. wrote the paper with input from all authors. 

\section*{Competing Interests }
The authors declare no competing interests.

\clearpage

\section*{Figure Legends/Captions } 

\clearpage


\section*{Methods}
\subsection*{Density matrix renormalization group}
DMRG\cite{WhitePRL1992,WhitePRB1993} is employed to solve $n$-MMHK in one dimension and a quasi-2d ladder. Eq.~\ref{ohk} can be expressed as a sum over $\mathbf{k}$ for the $n$-site Hubbard model with a complex hopping matrix $g_{\alpha \alpha'} (\mathbf k)$. DMRG is then performed for each of these Hubbard systems (for each value of $\mathbf{k}$) and the MMHK ground state energy is computed by averging over the ground state energies of each Hubbard cluster.
\subsection*{Exact diagonalization}
ED with Lanczos algorithm\cite{Lanczos1950} is used to determine the ground state of $16$-MMHK model at zero temperature and to compute the dynamical quantities including the spectral function and spin susceptibility. For this model, we need to solve the $16$-site Hubbard cluster for each $\mathbf{k}$ ($400$ clusters in total). For each $16$-site Hubbard cluster, we use Lanczos iteration to solve for the ground state at each total momentum ($16$ in total) with energy accuracy $\sim 10^{-12}$ and then locate the true ground state among them. Based on that state, we then conduct another Lanczos iteration to estimate the dynamical quantities. 

\section*{Data Availability}
The data and scripts needed to reproduce the figures are openly available in \url{10.5281/zenodo.17096693}.
\section*{Code Availability} 
The exact diagonalization code used for solving the $16$-MMHK model can be obtained at \url{https://github.com/wztzjhn/quantum_basis}. And the DMRG algorithm is implemented through TeNPy which can be accessed at  \url{https://tenpy.readthedocs.io/en/v1.0.6/reference/tenpy.algorithms.dmrg.html}. All other codes will be made available upon request. \\
\\

\begingroup
\renewcommand{\refname}{Methods-only references} 

\endgroup

\newpage
\section*{Extended data}

\begin{extfigure*}[h!]
	\centering
	\includegraphics[width=0.7\textwidth]{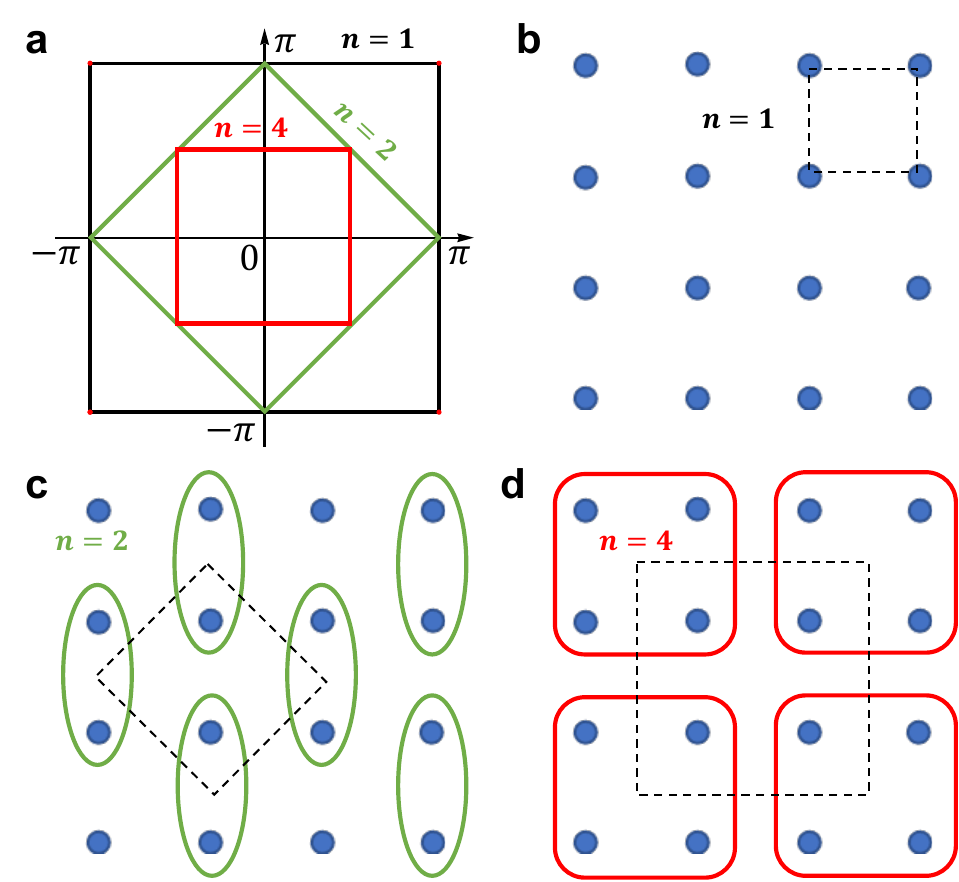}
	\caption{{\bf Scheme of $n$-MMHK in two dimensions.} {\bf a} is reproduced from Fig. 1{\bf b} for convenience. Alternative real-space construction by grouping the atoms into cells (encircled by green/red lines) with an updated lattice constant. {\bf b}, Original unit cell in the band HK model with primitive unit cell in dashed line. {\bf c}, The $2$-MMHK model leads to a doubling of the unit cell size and the lattice constant (dashed line) {\bf d}, The $4$-MMHK model results in a quadrupling of the unit cell size.  When $n= N $, the unit cell contains all the sites.}
	\label{fig:ex_rbz}
\end{extfigure*}

\begin{extfigure*}[h!]
	\centering
	\includegraphics[width=\textwidth]{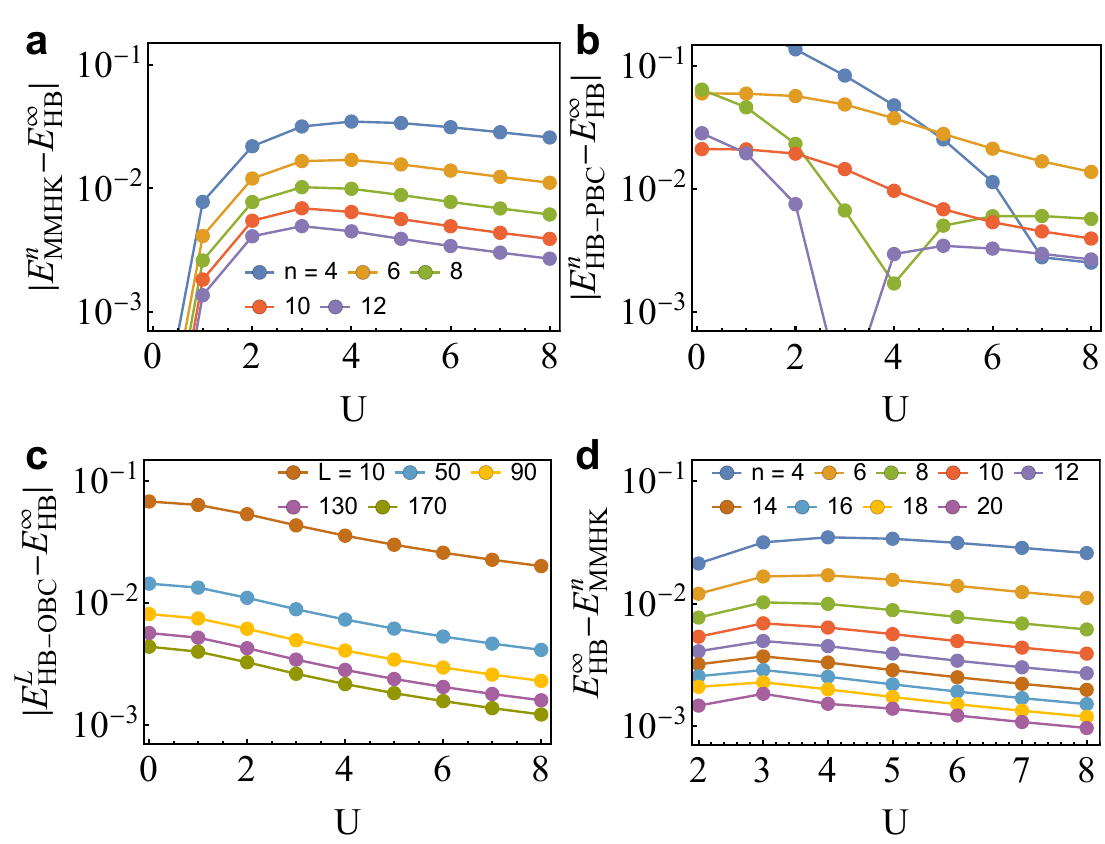}
	\caption{{\bf Ground-state energy accuracy of $n$-MMHK and finite-size Hubbard clusters in one dimension.} Comparison of the deviation in ground-state energy relative to the Bethe ansatz result for the TD limit ($E_{\rm HB}^\infty$) among different model settings. {\bf a} shows the deviation of the $n$-MMHK model ($E_{\rm MMHK}^n)$, which is consistently less than $1\%$ for $n\geq 10$. {\bf b} gives the deviation of a periodic $n$-site Hubbard chain ($E_{\rm HB-PBC}^n$), solved by ED, with the same legend as {\bf a}. The scattered $U$-dependence for various $n$ indicates non-systematic finite-size effects. {\bf c} displays the deviation of an open Hubbard chain with length $L$ ($E_{\rm HB-OBC}^L$), obtained from standard DMRG. {\bf d} is similar to {\bf a} but with larger $n$ (the number of mixed momenta) and the $n$-MMHK models are solved by DMRG for $n>12$. Comparing {\bf c} and {\bf d}, we find that $n$-MMHK exhibits more stable accuracy across the range of $U$, while open Hubbard chains show larger errors for small $U$.}
	\label{fig:1dgsen}
\end{extfigure*}

\begin{extfigure*}[h!]
	\centering
	\includegraphics[width=0.5\textwidth]{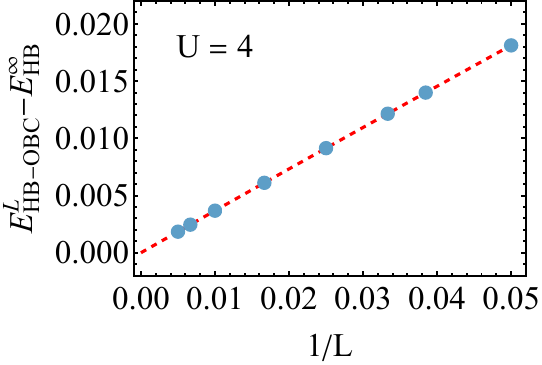}
	\caption{ {\bf Scaling of ground-state energy accuracy in an open Hubbard chain.} Standard DMRG results for an open finite Hubbard chain at $U=4$ are shown as deviations from the Bethe–ansatz ground-state energy. The deviation scales linearly with $1/L$.}
	\label{fig:ex_1dfitting}
\end{extfigure*}

\begin{extfigure*}[h!]
	\centering
	\includegraphics[width=\textwidth]{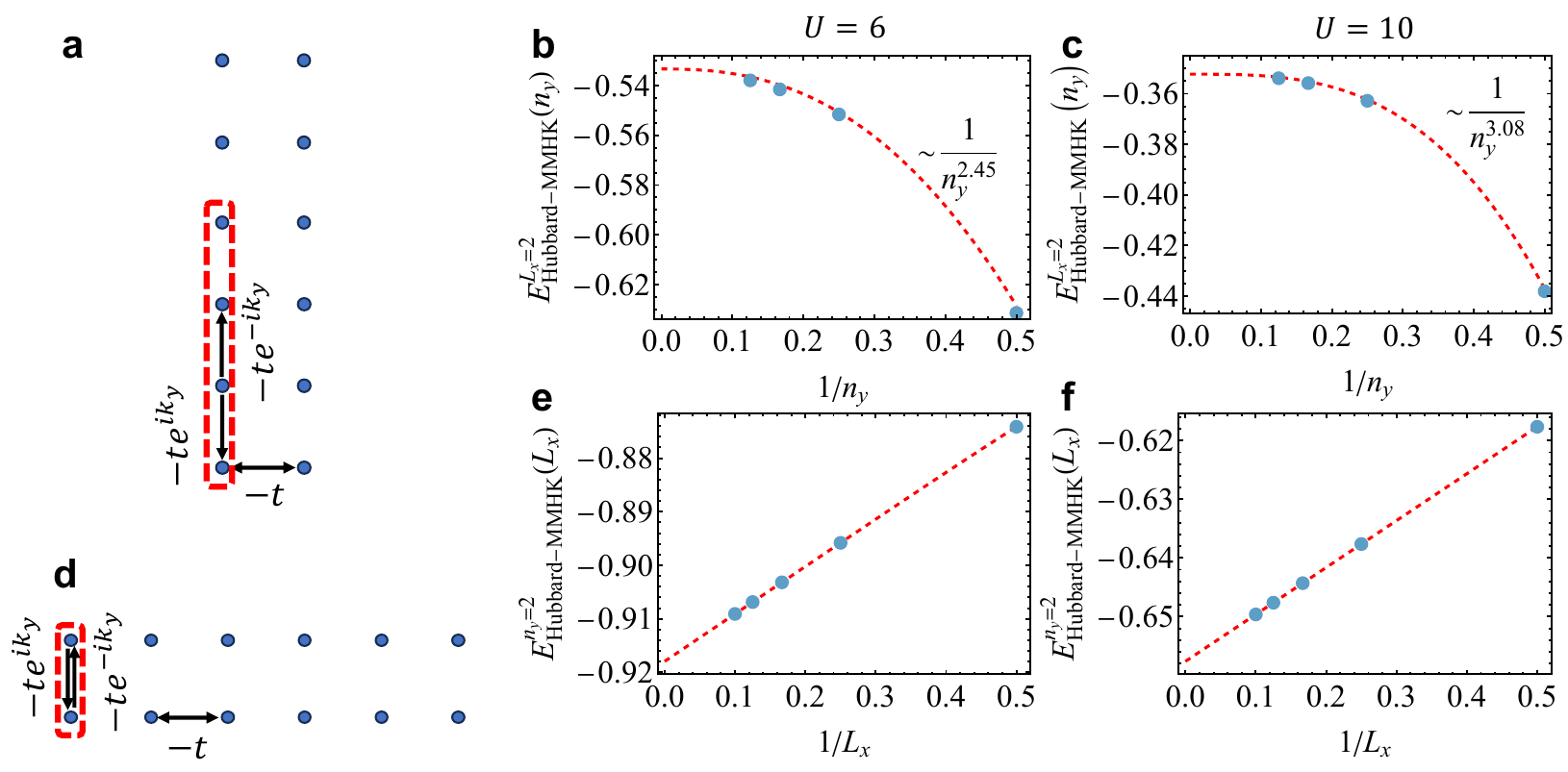}
	\caption{ {\bf Two-leg Hubbard–MMHK hybrid ladder.} The hybrid ladder consists of MMHK sites arranged along the $y$-direction and Hubbard sites with open boundary conditions applied along the $x$-direction. The system is extended along ({\bf a}) the $y$-direction and ({\bf d}) the $x$-direction. The ground-state energy of the hybrid ladder is shown for $U=6$ (smaller than bandwidth $w=8$) in ({\bf b, e}) and for $U=10$ (larger than $w$) in ({\bf c, f}). {\bf b, c} display the ground-state energy as a function of $n_y$, the number of MMHK sites along $y$-direction, corresponding to the setup in ({\bf a}). {\bf e, f} show the energy as a function of $L_x$, the length along the $x$-direction, corresponding to the setup in ({\bf d}). These results are all obtained using DMRG and the dashed lines represent the fitting to the scaling behaviors. {\bf b, c}, The energy converges faster than the $1/n^2$ behavior observed in the purely one-dimensional $n$-MMHK cases. {\bf e, f}, The energy converges linearly as $L_x$ increases.
    }
	\label{fig:2dladder}
\end{extfigure*}

\begin{extfigure*}[h!]
	\centering
	\includegraphics[width=\textwidth]{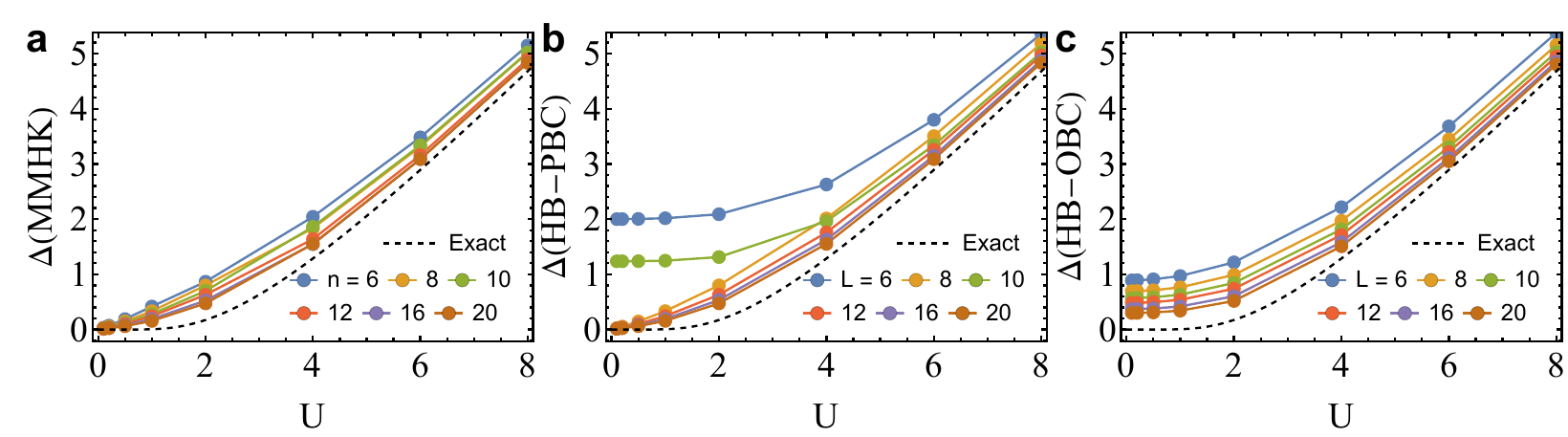}
	\caption{{\bf Single-particle charge gap for $n$-MMHK, periodic, and open Hubbard chains in one dimension. } The single-particle charge gaps are shown for ({\bf a}) the $n$-MMHK model, ({\bf b}) a periodic Hubbard chain with length $L$, and ({\bf c}) an open Hubbard chain with length $L$, all obtained from DMRG simulations. In all panels, the dashed line denotes the exact result from Bethe ansatz. {\bf a}, $n$-MMHK captures the vanishing gap at $U=0$ (even for small $n$) and slowly converges to the exponential singularity (Bethe ansatz) at small $U$ as $n$ increases. {\bf b}, Periodic Hubbard chain exhibits non-systematic $L$-dependence, with large $L$ behaviors close to MMHK. {\bf c}, Open Hubbard chain captures the larger-$U$ regime accurately but slowly converges to the correct $U=0$ limit due to finite length. 
    }
	\label{fig:ex_1dgap}
\end{extfigure*}

\begin{extfigure*}[h]
	\centering
	\includegraphics[width=\textwidth]{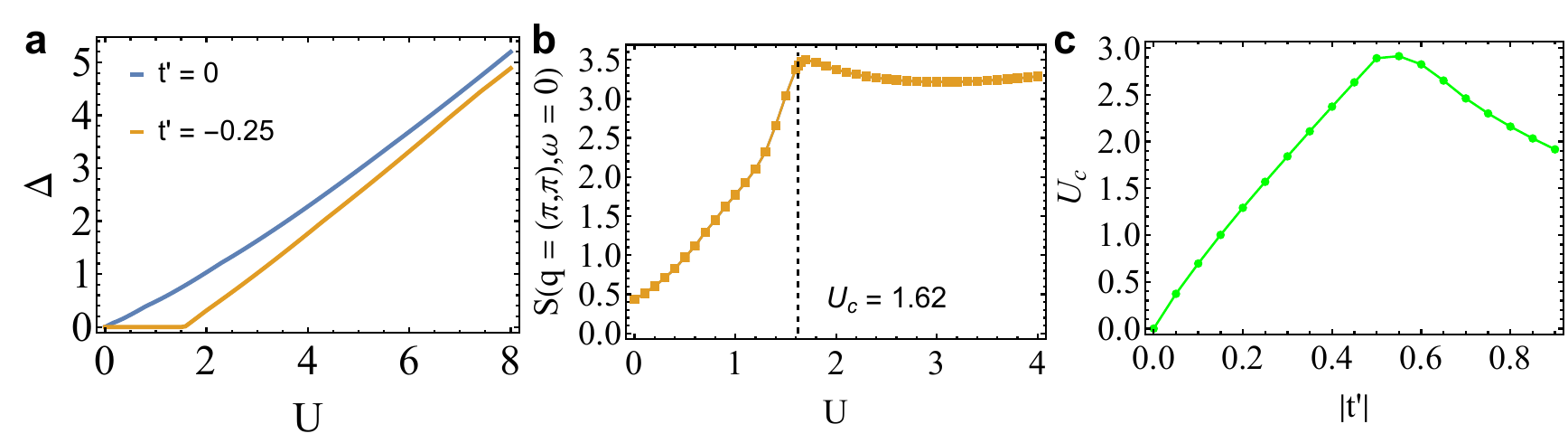}
	\caption{{\bf Mott transition in the $4$-MMHK model.} {\bf a}, Evolution of single-particle charge gap $\Delta$ with $U$ at half-filling for $t'=0$ and $t'/t=-0.25$. At $t'=0$, the gap opens for any finite $U$, while in the frustrated case ($t'/t=-0.25$), the gap opens only for $U>U_c=1.62$, indicating a Mott transition. {\bf b}, Antiferromagnetic spin susceptibility at $t'=-0.25$ and $\beta=100$ as a function of $U$, showing a cusp at $U_c$. (c) The critical interaction strength $U_c$ for the Mott transition as a function $\big| t' \big|$(the sign of $t'$ does not make a difference in $U_c$). Initially, $U_c$ grows almost linearly with $|t'|$. It saturates around $|t'|\approx0.55$ and then decreases with further increasing $|t'|$.}
	\label{fig:ex_2dgap}
\end{extfigure*}

\begin{extfigure*}[h]
	\centering
	\includegraphics[width=\textwidth]{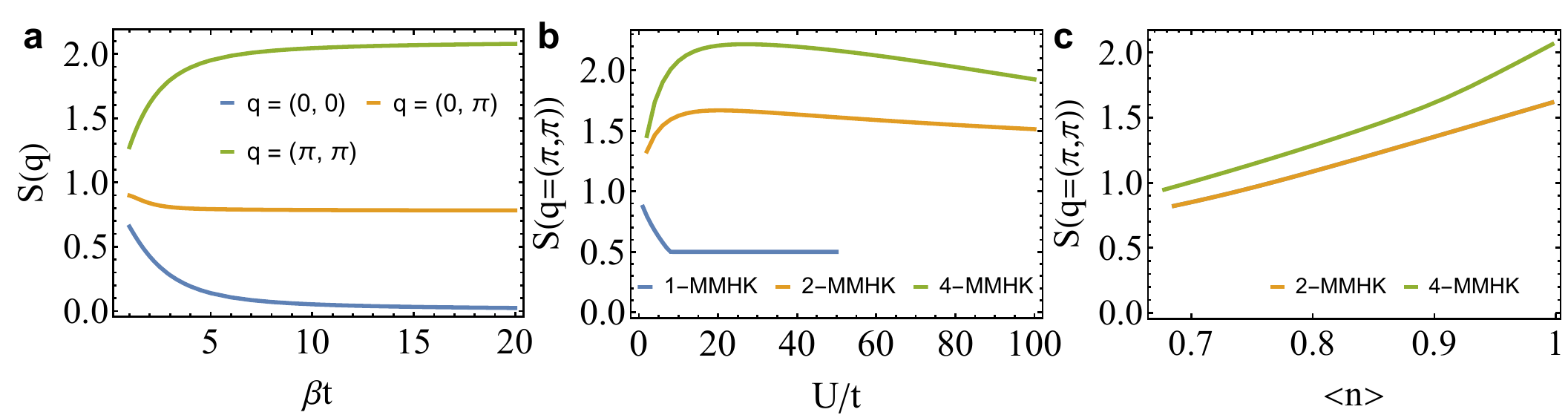}
	\caption{{\bf Equal-time spin correlation for $n$-MMHK model.} {\bf a}, Equal-time spin correlation as a function of inverse temperature for different transfer momenta $q$ for $4$-MMHK at $U=10$. The $q=(\pi,\pi)$ case dominates the spin correlation, indicating the leading magnetic fluctuation is of the antiferromagnetic type. {\bf b}, Equal-time antiferromagnetic correlation as a function of $U$ for different $n$-MMHK models. As the number of mixed momenta $n$ increases from $1$ (band HK), the antiferromagnetic correlation strives, indicating a qualitatively better simulation of Hubbard physics. {\bf c}, Equal-time antiferromagnetic correlation as a function of density at $U=10, \beta t=20$ for $2$- and $4$-MMHK models. The antiferromagnetic correlation remains for a wide range of doping, essential for capturing the more complicated doping physics in the Hubbard model. }
	\label{fig:ex_spindiffq}
\end{extfigure*}

\begin{extfigure*}[h!]
	\centering
	\includegraphics[width=0.8\textwidth]{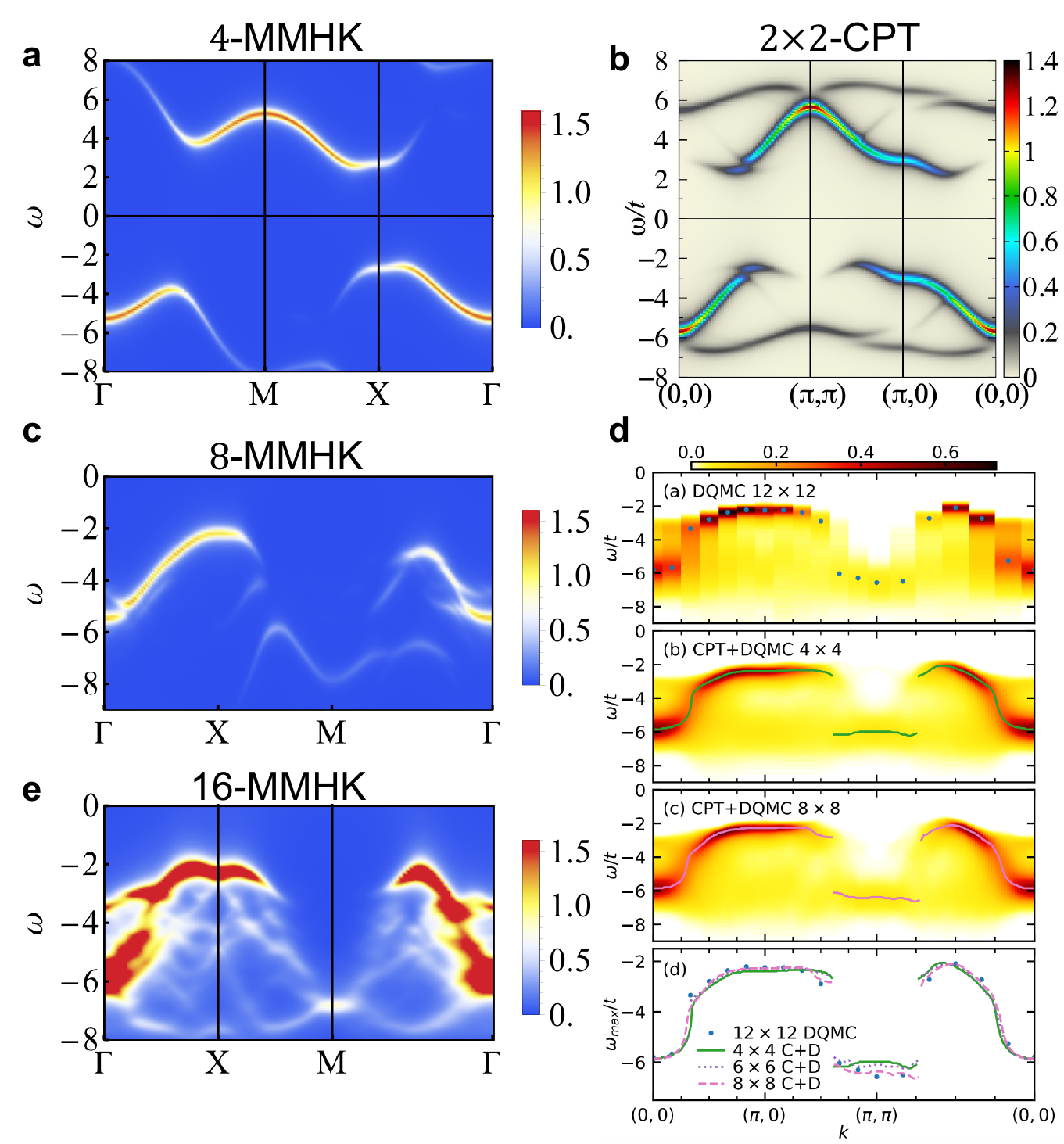}
	\caption{{\bf Spectral functions of $n$-MMHK models with various $n$.} We compare the half-filled ($U=8$) spectral function $A({{\bf k}},\omega)$ among $n$-MMHK models (at zero temperature with a Lorentzian broadening of $0.2$) and other methods for solving the Hubbard model. ({\bf a}) $4$-MMHK (or $2\times2$-MMHK model) and ({\bf b}) $2\times2$ cluster perturbation theory (reprinted from \refdisp{Sekiprb2016}) show qualitatively similar leading features, including peak positions and gap size. The ({\bf c}) $8$-MMHK and ({\bf e}) $16$-MMHK results progressively improve the description of weaker sub-leading features, becoming quantitatively comparable to ({\bf d}) state-of-the-art determinant quantum Monte-carlo and/or cluster perturbation theory (reprinted from \refdisp{Huangprr2022}) at $\beta=16$. This underscores that as $n\rightarrow N$, MMHK becomes the Hubbard model.}
	\label{fig:ex_Akw}
\end{extfigure*}

\begin{extfigure*}[h!]
	\centering
	\includegraphics[width=0.6\textwidth]{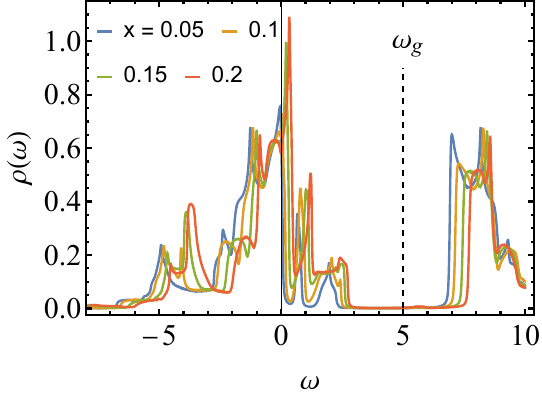}
	\caption{{\bf Density of states of the hole-doped $4$-MMHK model and the choice of $\omega_g$.} We present the density of states at varying hole-doped densities $x$ with $U=10,\beta=30$. The cutoff $\omega_g$ is chosen at the gap for calculating the low energy spectral weight with Eq.~(10).}
	\label{fig:ex_cutoff}
\end{extfigure*}

\begin{extfigure*}[h!]
	\centering
	\includegraphics[width=\textwidth]{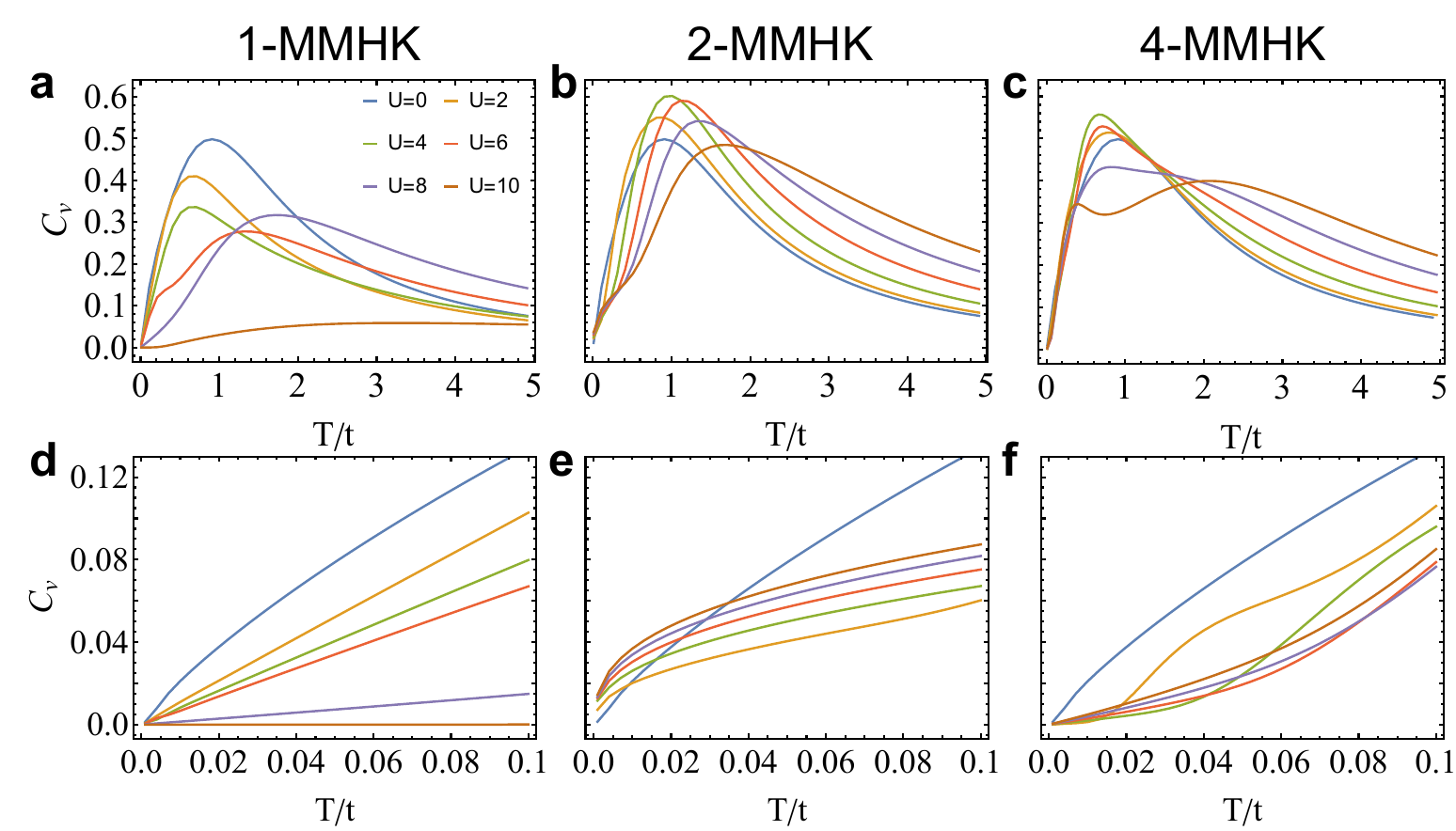}
	\caption{{\bf Heat capacity for $n$-MMHK models with $n=1,2,4$ at various $U$.} {\bf a-c} are for high temperatures, while {\bf d-f} are for low temperatures.}
	\label{fig:ex_HC}
\end{extfigure*}

\end{document}